\documentclass[]{aa}  %[referee]

\usepackage{graphicx}
\usepackage{color}
\usepackage{fixltx2e}
\usepackage{txfonts,booktabs}
\begin{document}

\title{A new HW Vir binary from the Palomar Transient Factory}

\subtitle{PTF1\,J072455.75$+$125300.3 - An eclipsing subdwarf B binary with a M-star companion}
\titlerunning{PTF1\,J072456$+$125301 - An eclipsing subdwarf B binary with an M-star companion}
\authorrunning{M. Schindewolf et al.}	
\author{M. Schindewolf \inst{1}\and D. Levitan \inst{2} \and U. Heber \inst{1} \and H. Drechsel \inst{1} \and V. Schaffenroth \inst{1,3}  \and T.  Kupfer \inst{4} \and T. Prince \inst{2}}  

 \institute{Dr.\,Remeis-Observatory \& ECAP, Astronomical Institute, Friedrich-Alexander University Erlangen-N\"urnberg, Sternwartstr.~7, 96049
 Bamberg, Germany\\
              \email{markus.schindewolf@fau.de}
	\and
	Division of Physics, Mathematics and Astronomy, California Institute of Technology, Pasadena, CA 91125, USA
         \and
         Institute for Astro- and Particle Physics, University of Innsbruck, Technikerstr. 25/8, 6020 Innsbruck, Austria
         \and
         Department of Astrophysics/IMAPP, Radboud University Nijmegen, P.O. Box
         9010, 6500 GL Nijmegen, The Netherlands
             }

   \date{Received.../ Accepted ...}

 \abstract{
We report the discovery of an eclipsing binary -- PTF1\,J072456$+$125301-- composed of a subdwarf B (sdB) star ($g'=17.2^m$) with a faint companion. 
Subdwarf B stars are core helium-burning stars, which can be found on the extreme horizontal branch. About half of them reside in close binary systems, but few are known to be eclipsing, for which fundamental stellar parameters can be derived.\newline
We conducted an analysis of photometric data and spectra from the Palomar 60'' and the 200" Hale telescope respectively. A quantitative spectral analysis found an effective temperature of $T_{\text{eff}}=33900\pm350$\,K, log g = $5.74\pm0.08$ and log($n_{\text{He}}/n_{\text{H}}) = -2.02 \pm0.07$, typical for an sdB star. The companion does not contribute to the optical light of the system, except through a distinct reflection effect.
From the light curve an orbital period of 0.09980(25)\,d  and a system inclination of $83.56\pm0.30\,^{\circ}$ were derived. The radial velocity curve yielded an  orbital semi-amplitude of $K_1=95.8\pm 8.1\,\text{km s$^{-1}$}$. The mass for the M-type dwarf companion is $0.155\pm0.020\,M_{\sun}$. PTF1\,J072456$+$125301 has similar atmospheric parameters to those of pulsating sdB stars (V346 Hya stars). Therefore it could be a high-priority object for asteroseismology, if pulsations were detected such as in the enigmatic case of NY Vir. 
}


\maketitle
%
%________________________________________________________________
\section{Introduction}
Hot subdwarf stars (sdBs and sdOs) are core helium-burning stars with thin hydrogen envelopes that are unable to sustain hydrogen shell burning \cite{heber09}. This type of star dominates the population of faint blue stars at high galactic latitudes and can be found both in the old disk as well as in the halo \cite{ferraro}. With regard to galaxy evolution, sdB stars play an important role because they are believed to be a dominating source for the "UV upturn phenomenon" that can be observed in elliptical galaxies and galaxy bulges \cite{brown97}.\newline
Subluminous B stars are identified with evolutionary models for Extreme Horizontal Branch (EHB) stars.
To form an sdB star, the progenitor star has to lose almost all of its hydrogen envelope rather quickly, right at the onset of helium burning. The remaining thin part of the envelope is inert. Close binary evolution is very likely to be responsible for the mass loss on the red giant branch (RGB) that is needed to form an sdB star \cite{maxted01}. A significant fraction of sdB stars have composite spectra indicating the presence of a main sequence companion of F, G or K type (Ferguson et al. 1984, Stark \& Wade 2003). Such binaries have long periods of several hundreds of days to more than 1000 \cite{vos}. About 50\,\% of the known single-lined sdB stars reside in close binaries with periods less then 30 days, hosting invisible companions that are either white dwarfs or very low mass dwarfs (Maxted et al. 2001, Copperwheat et al. 2011). To understand the formation of the sdB binaries, two possible channels have been proposed, both requiring interacting binary systems (Han et al. 2002, Han et al. 2003):

(i) The first one is called the \grqq stable Roche lobe overflow (RLOF) channel\grqq.  Stable mass transfer is taking place, originating from a low mass giant filling its Roche lobe close to the tip of the RGB. During this time it loses most of its envelope due to the Roche lobe overflow. When the hydrogen envelope has been reduced sufficiently, the mass transfer stops and the star begins to shrink. If the degenerate core is massive enough, a helium flash is possible. Subdwarf B systems originating from the RLOF channel are predicted to have long periods from 10\,d up to 1000\,d.  
\newline
(ii) The second one is the \grqq common envelope ejection channel\grqq \,(CEEC). Again mass transfer occurs close to the tip of the red giant branch of the sdB progenitor, but the mass transfer rate is so high that a common envelope (CE) is formed. Due to friction in the envelope, both stars approach each other. This leads to a release of orbital energy and angular momentum which is used to eject the envelope. This process results in a close binary system. If the core of the initially more massive star is still able to ignite helium burning, it becomes an sdB. The result would be an sdB star with a main sequence companion in a short period binary such as the HW~Vir star discussed in this paper.
If a second common envelope phase occurs the companion would be a white dwarf in a close orbit to the sdB star.\newline

The mass distribution of sdB stars is found to peak at the canonical sdB mass of $0.47\,M_{\sun}$  \cite{dorman}. Observational mass determinations for a few objects are consistent with such predictions ($0.47\pm0.031\,M_{\odot}$ (Van Grootel et al. 2014, Fontaine et al. 2012).
\newline
Subdwarf B stars in close binary systems with low-mass dwarf companion have been known for decades \cite{kilkenny}. Those companions have been discovered by means of the reflection effect in their light curves. If those systems are eclipsing, they are called HW Vir systems, named after the first discovered system \cite{menzies}. They mostly have short periods around 0.1\,days and a late M-type companion (which can usually only be detected by its reflection effect, caused by the high temperature difference of both components). The short period indicates that they have undergone a common envelope phase followed by a spiral-in phase. Up to now, only fifteen such systems are known \cite{schaffenroth14a,schaffenroth14b}).\newline
The canonical mass remains valid for HW Vir systems, although being the result of binary evolution \cite{han03}.
 \newline
Because such eclipsing binaries offer the opportunity to determine the stellar masses, detailed analyses are crucial for as many such objects as possible. 
\newline
In this paper, we present observations and analysis of a new HW Vir star discovered by the Palomar Transient Factory (PTF). The PTF is a wide field survey that searches for variable sources and optical transients.
It makes use of the 1.2\,m Oschin Telescope at Palomar Observatory, equipped with eleven 12k $\times$ 8k CCD cameras, and the Palomar 60'' telescope \cite{law}. With each exposure 7.2 square-degrees are covered. In a typical night  200 square-degrees or more are observed \citep{levitan11}.
\newline
We describe the spectroscopic and photometric observations in Sect. 2 and their analysis in Sect. 3 (spectroscopy) and Sect. 4 (photometry). In Sect. 5 we give a summary of the results from the analysis.

\section{Observations}
Spectroscopic and photometric data presented here were obtained at the Palomar 60”  (hereafter P60) \cite{cenko06}  and Palomar 200” (hereafter P200) telescopes, respectively. 
	\subsection{Spectroscopy}
The spectra of PTF1\,J072456$+$125301 (PTF0724 hereafter) were taken at the Palomar 200'' telescope with the Double Spectrograph \grqq DBSP\grqq{} \cite{oke82} using a low resolution (R$\sim$ 1500). The observations took place on August 10, 2010 and were scheduled to cover the entire orbit at equally distributed phases. Thirty eight spectra were collected, covering a range fom 3000\,{\AA} to 11000\,{\AA} . The exposure time was set between 600\,s and 900\,s per spectrum to prevent orbital smearing. A grating of 600 lines/4000\,mm-blaze was used at a grating angle of $27^{\circ}17''$. All P200 spectra were reduced using optimal extraction \cite{horne86} as implemented in the Pamela Code \cite{marsh89}, as well as the Starlink packages kappa, figaro and convert. 
The spectrum is readily classified as that of an sdB star, according to the strong Stark-broadened Balmer and weak helium lines.

	\subsection{Photometry}
We obtained a total of 1,162 photometric observations of PTF0724 with the P60 over 7 nights (details of each night are in Table A.2 in the appendix. All observations were made using a g' filter and with an exposure time of 45 sec with a dead time of approximately 21 sec between exposures. \newline
The P60 photometric observations were de-biased, flat-fielded, and astrometrically calibrated using the P60 pipeline \cite{cenko06}. Stars in each individual image were identified using the Sextractor Package \cite{Bertin} and instrumental magnitudes were measured using optimal point-spread function photometry \cite{naylor} as implemented by the Starlink package Autophotom\footnote{ The Starlink Software Group homepage can be found at \url{http://starlink.jach.hawaii.edu/starlink} }.

Calibrated light curves were calculated using the relative photometry algorithm described in the appendix of Ofek et al. (2011) and in Section 3.1 of Levitan et al. (2011). Briefly, this algorithm identifies non-varying (over the course of the observations) “calibration” stars in the field of view and minimizes the scatter of their instrumental magnitudes by adjusting the zero-point of each individual exposure. We simultaneously fit to a reference catalog (here, USNO-B 1.0) to find an absolute calibration. This minimization is calculated using an iterative, matrix-based, least-squares approach and the resulting zero-points can then be applied to the target of interest. We note that we did not utilize the de-trending features of the algorithm as described in Levitan et al. (2011) since only one telescope/instrument was used. Measurement uncertainties are calculated from a combination of the Poisson error and the fit error. However, we note that this estimate does not include systematic errors and they are thus likely under-estimated. Calculating the systematic error accurately can be done by finding the minimum scatter observed for bright stars on each exposure, but doing so requires substantially more sources than are present in the small field of view here. \newline
The lightcurve is characterized by a strong reflection effect and primary as well as secondary eclipses. This, together with the spectral properties of the star (see 3.2), classifies PTF0724 as an HW Vir star.\newline
The primary minimum is quite deep, with the flux descending to about 25\% of the flux at phase 0.25 ($g'=17.2^m$ at this phasepoint). For further analysis the lightcurve was normalized to 1 at phase 0.25. The ephemeris was determined by fitting Gaussians to the primary eclipses via the standard Marquardt-Levenberg algorithm \cite{levenberg}. The period could be easily derived, because several lightcurves covered more than one full orbit, containing two primary minima.\newline
The ephemeris of the primary minimum was found to be:\newline
HJD$=2455295.64113(8) + 0.09980(25)\cdot E$\newline
(with the eclipse number E).
%______________________________________________________________

\section{Spectroscopic analysis}
\subsection{Radial velocity curve}
\begin{figure}
\centering
\includegraphics[width=\linewidth]{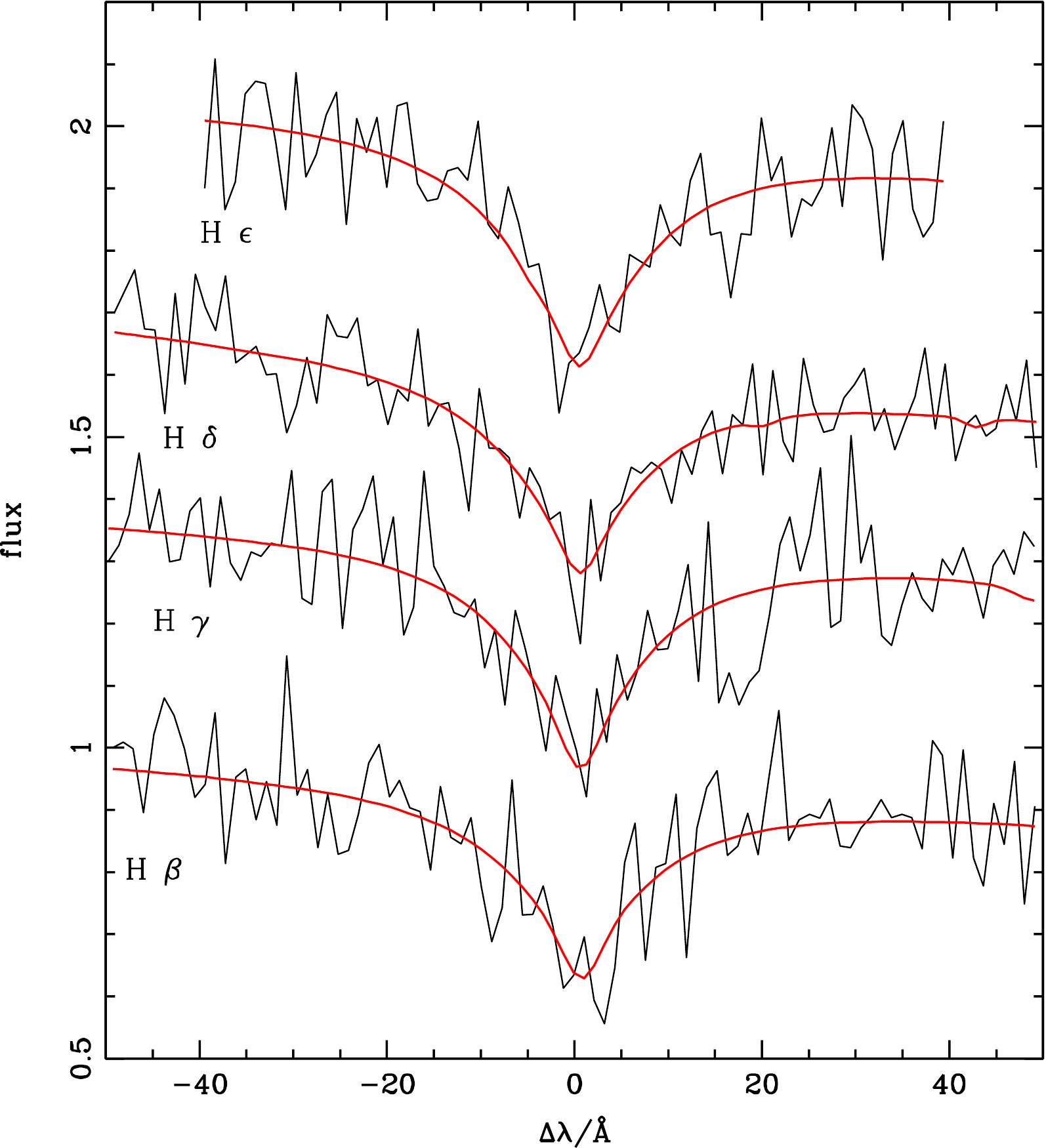}
\caption{Example fit to several Balmer lines of an individual spectrum, to determine the radial velocities.}
\label{lines}
\end{figure}

We begin by measuring and plotting the radial velocity (RV) curve. Since the system is single-lined, the analysis is straight forward. We made use of SPAS, the Spectrum Plotting and Analysis Suite \cite{hirsch} to determine the radial velocity for each blue spectrum (see table A1). The RVs were measured by fitting a set of mathematical functions (Gaussians, Lorentzians and polynomials) to the spectral lines (see Fig. \ref{lines}). Three functions are used to match the continuum (polynomials), the line (Lorentzians) and the line core (Gaussians), respectively and mimic the typical Voigt profile of spectral lines. The profiles are fitted to all suitable lines simultaneously using $\chi^2$ minimization and the RV shift with respect to the rest wavelengths with the associated 1$\sigma$ error is measured.

\begin{figure}[h!]
\centering
\includegraphics[width=\linewidth]{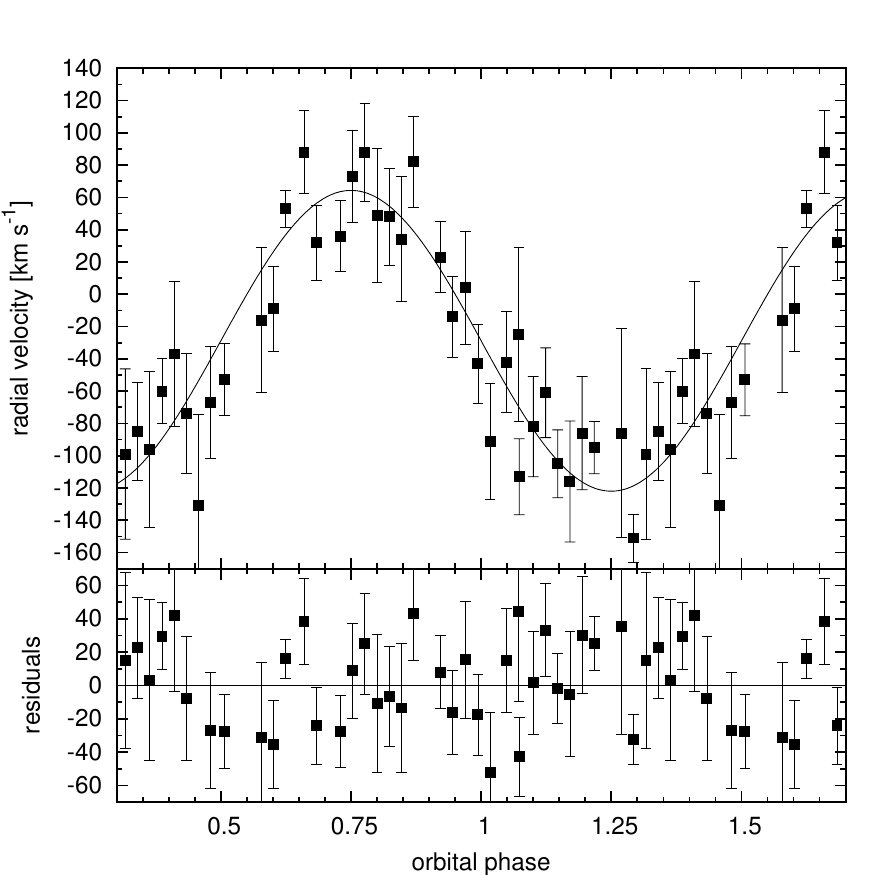}
\caption{Radial velocity plotted against orbital phase of PTF0724. In the bottom panel the residuals can be seen.}
\label{rv}
\end{figure}

Because we find the phase shift between the primary and secondary minima is $0.5009\pm0.001$, the eccentricity of the  orbit of PTF0724 must be very small (Fig. \ref{lc}). Therefore we fitted a sine curve to the RVs. Fig. \ref{rv} shows the phased RV curve with the best-fit solution. All phases of the orbit are well covered. We derived a semiamplitude of K = $95.7 \pm 8.1\,\text{km s$^{-1}$}$ and a system velocity of $\gamma = -26.7\pm 5.5\,\text{km s$^{-1}$}$ with a bootstrapping approach using 1200 iterations. Thereby the observational data is resampled randomly and fitted again. The standard deviation of a parameter is regarded the 1-$\sigma$ error of the parameter. To estimate the contribution of systematic effects to the error budget, we modified the $\chi^2$ of the best solution by adding systematic errors in quadrature until the reduced $\chi^2$ reached $\sim$1.0. \newline 
The period derived from the sinusoidal fit (P=$ 0.0976\pm0.0042$\,d) is consistent with the period derived from photometry (P=0.09980(25)\,d) resulting in a mass function of f(m)=0.0093$\pm$0.002 M$_{\odot}$.The inclination of the orbit is well constrained by photometry $(i=83.56\pm0.30^{\circ}$, see Sect. 4).  Adopting the canonical mass of 0.47\,M$_{\odot}$ we find a mass ratio of q=$0.327\pm0.033$ by making use of the mass function.

\subsection{Atmospheric parameters}
Due to the poor signal-to-noise ratio (S/N$\sim10$) of the individual spectra, we shifted and coadded the individual velocity-corrected spectra to improve the S/N ratio to be able to derive reliable atmospheric parameters.\newline
Effective temperature, surface gravity and helium abundance were determined by fitting synthetic spectra to the Balmer and helium lines of the coadded spectra. The synthetic spectra were calculated by using LTE model atmospheres with solar metallicity and metal line blanketing\cite{heber2000}.
The derived parameters are: T$_{\text{eff}} = 33900\pm 350\,\mbox{K}$, log g = $5.74 \pm 0.08$ and log y = $-2.02 \pm 0.07$.\newline
The errors of the parameters were calculated with a bootstrapping approach.
\newline

\begin{figure}[h!]
\centering
\includegraphics[width=\linewidth]{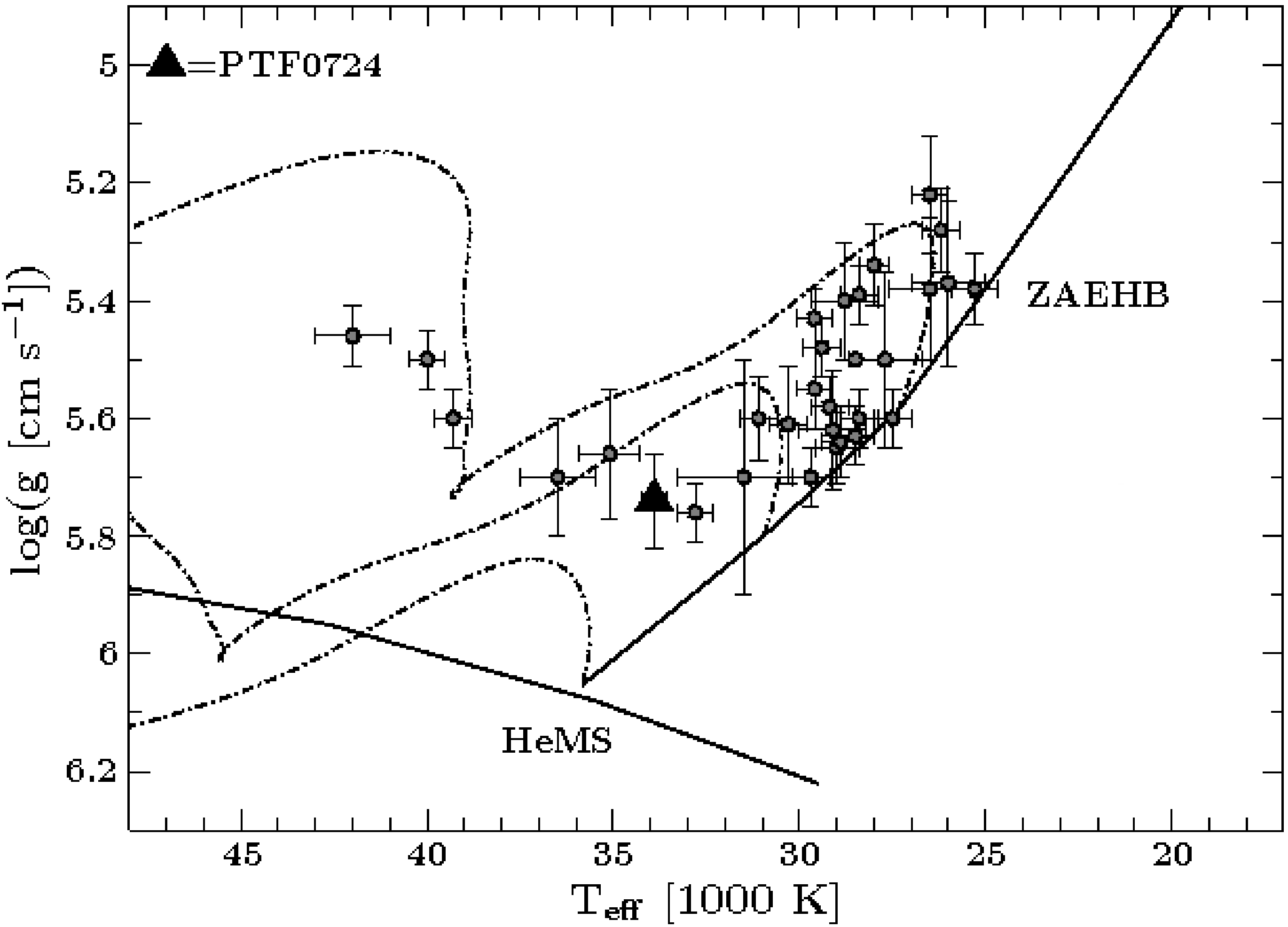}
\caption{Position of PTF0724 (black triangle) in a ($T_{\text{eff}}$, log g) diagram in comparison the known sdB+dM systems in close binary systems with known periods (grey circles, \citealt{kupfer14}). Overplotted  EHB evolutionary tracks by Han et al. (2002) using $0.50$\,M$_\odot$  models (dashed-dotted lines). In addition the Zero-Age Horizontal Branch (ZAEHB) as well as the helium main sequence (HeMS) are shown.  }
\label{evolution}
\end{figure}

In Fig. \ref{evolution} the position of PTF0724 in a ($T_{\text{eff}}$, log g) diagram is compared to the known sdB+dM systems in close binary systems with known periods \cite{kupfer14}. 
Its position in the graph indicates that the primary component in PTF0724 is a normal sdB close to core helium exhaustion. Similar systems lie in the same regime. There are two classes of multi-periodic pulsating stars known among the sdB stars, which show either g-modes (361~Hya stars) or p-modes (V1093~Her stars). 
It is worthwhile to mention that its atmospheric parameters place PTF0724 in a regime where the driving of g-mode pulsations is predicted to be largest \citep{charpinet07}. 
%______________________________________________________________

\section{Lightcurve analysis}
\begin{figure}[h!]
\centering
\includegraphics[width=\linewidth]{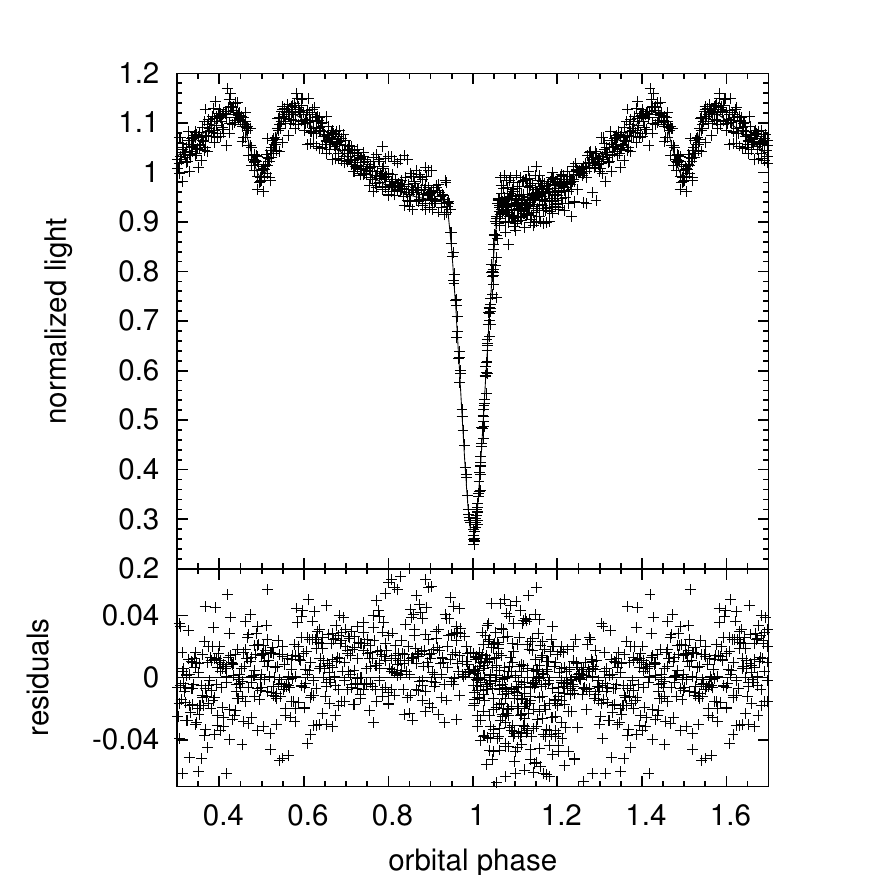}
\caption{Phased lightcurve of PTF0724. The solid line represents the best fitting model for q=0.33. In the bottom panel the residuals can be seen. Some outlying datapoints, mainly before and after the primary eclipse, are visible. This was most likely caused by the changing airmass. As the exact coverage of the depth and form of the bottom of the primary eclipse is crucial to get correct results,  the datapoints around phase 1 were weighted higher, resulting in smaller residuals. }
\label{lc}
\end{figure}

The seven lightcurves, derived from the observations in Table 2, were phased, normalized and merged into one lightcurve (see Fig.\ref{lc}).  Its analysis proceeds as described in Drechsel et al. (2001) , for details see also Schaffenroth et al. (2013).

\begin{table}[h!]
\caption{Model parameters of the lightcurve solution for q=0.33}
\label{ptf33}
\begin{tabular}{lcl}
\hline
\noalign{\smallskip}
\multicolumn{3}{l}{Fixed parameters:}\\
\noalign{\smallskip}
\hline
\noalign{\smallskip}
%$q\,(=M_{2}/M_{1})$ & & $0.33$\\
$T_{\rm eff}(1)$&[K]&\multicolumn{1}{l}{33900}\\
$g_1^b$&&\multicolumn{1}{l}{1.0}\\
$g_2^b$&&\multicolumn{1}{l}{0.32}\\
$x_1^c$&&\multicolumn{1}{l}{0.190}\\
$A_1^a$&&\multicolumn{1}{l}{1.0}\\
$\delta_2^d$&&\multicolumn{1}{l}{0.0}\\
$l_3^e$&&\multicolumn{1}{l}{$0.00$}\\
\noalign{\smallskip}
\hline
\noalign{\smallskip}
\multicolumn{3}{l}{Fitted parameters:}\\
\noalign{\smallskip}
\hline
\noalign{\smallskip}
$i$ & [$^{\rm \circ}$] & $83.56 \pm 0.3$\\
$T_{\rm eff}(2)$ & [K]& $3300 \pm 300$\\
$A_2^a$ & & $1.8 \pm 0.2$\\
$\Omega_1^f$&&$5.544 \pm 0.20$\\
$\Omega_2^f$&&$2.885 \pm 0.15$\\
$x_2^c$&&$0.510 \pm 0.05$\\
$\left(\frac{L_1}{L_1+L_2}\right)^{\textbf{g}}$&&$(0.99978\pm 0.00064) $\\
$(\delta_1^{\textbf{d}})$&&$(0.0006 \pm 0.0003$)\\

\noalign{\smallskip}
\hline
\noalign{\smallskip}
\multicolumn{3}{l}{Roche radii$^h$:}\\
\noalign{\smallskip}
\hline
\noalign{\smallskip}
$r_1$(pole)&[a]&$0.191 \pm 0.011 $\\
$r_1$(point)&[a]&$0.193 \pm 0.012 $\\
$r_1$(side)&[a]&$0.192 \pm 0.011 $\\
$r_1$(back)&[a]&$0.193 \pm 0.010 $\\
\noalign{\smallskip}
$r_2$(pole)&[a]&$0.209 \pm 0.013$\\
$r_2$(point)&[a]&$0.227 \pm 0.013 $\\
$r_2$(side)&[a]&$0.214 \pm 0.017 $\\
$r_2$(back)&[a]&$0.223 \pm 0.015 $\\
\noalign{\smallskip}
\hline
\end{tabular}\\
\tablefoot{\\
$^{a}$ Bolometric albedo\\
$^{b}$ Gravitational darkening exponent\\
$^{c}$ Linear limb darkening coefficient; taken from Wade et al. (1085)\\
$^{d}$ Radiation pressure parameter, see Drechsel et al. (1995)\\
$^{e}$ Fraction of third light at maximum\\
$^{f}$ Roche potentials\\
$^{g}$ Relative luminosity; $L_2$ is not independently adjusted, but recomputed from $r_2$ and $T_{\rm eff}$(2)\\
$^{h}$ Fractional Roche radii in units of separation of mass centres}
\end{table}

\begin{table}[h!]
\caption{Adopted stellar parameters for the components of PTF0724.}
\label{mass33}
\centering
\begin{tabular}{c|c|c}
		\toprule
		\multicolumn{3}{c}{PTF1J072456$+$125301}\\ 
		\toprule
		i&$^\circ$&$83.56 \pm 0.30$\\
		 %$M_{\rm sdB}$ & [$M_{\rm \sun}$] & $0.526\pm0.05$\\
		 $M_{\rm comp}$ & [$M_{\rm \sun}$] & $0.155\pm0.020$\\
		 $a$ & [$R_{\rm \sun}$] & $0.766\pm 0.041$\\
		$R_{\rm sdB}$ & [$R_{\rm \sun}$]& $0.1488\pm 0.007$\\
		$R_{\rm comp}$ & [$R_{\rm \sun}$]&$0.165\pm 0.008$\\
		$\log{g}(\rm sdB,phot)$ & & $5.760\pm0.015$\\
	    $\log{g}(\rm sdB,spec)$ & & $5.75\pm0.08$\\
	    \bottomrule
\end{tabular}
\end{table}

The LC analysis was performed by using the MORO code \citep[MOdified ROche Program, see][]{drechsel95}. The program is based on the Wilson-Devinney approach \cite{wilson} but uses a modified Roche model that takes the radiative pressure of the stars into account. The optimization of parameters, is achieved by the \textit{simplex} algorithm.
\newline
One of the main problems with lightcurve analysis is the large number of free parameters. For a single lightcurve 17 parameters have to be adjusted but parameter correlations exist.
However, some of the parameters could be constrained from the spectroscopic analysis or on theoretical grounds, e.g. limb and gravity darkening parameters (see Table 1). Too many free parameters tend to generate under-determined solutions and there is no guarantee of uniqueness. \\In particular the mass ratio, one of the most important solution parameters, is strongly correlated with other parameters (i.e the Roche potentials $\Omega$) and causes a degeneracy of solutions. Therefore, in a first step, the mass ratio was fixed, while the other parameters were adjusted. Thereafter the mass ratio was changed by a certain increment and the iterative process started again.\newline
The effective temperature and the surface gravity of the primary component were derived from the spectroscopic analysis (see Sect. 3.2). Because of its early spectral type, the gravity darkening exponent can be fixed at $g_1 = 1$, as one would expect for radiative outer envelopes \cite{zeipel}. For the companion, we used $g_2 = 0.32$ \cite{lucy}, as expected for convective envelopes. The linear limb darkening coefficient for the primary was interpolated from the table of  Wade\& Rucinski (1985) and fixed at $x_1 = 0.190$. The radiation pressure parameter \cite{drechsel95} of the secondary was set to zero because of the low temperature of the secondary star. That means that the effects of the radiation pressure excerted by this star are negligible. All trial runs with different sets of parameters resulted in negligible third light. Hence there is no evidence for a third body in the system, and $l_3$ was set to zero. 
We constrain the mass ratio to the range 0.29 < q < 0.36, as indicated by the spectroscopic measurements. For each mass ratio a large grid of synthetic lightcurves was calculated to ensure that the final solution corresponds to the global minimum.
\newline
To estimate the quality of the lightcurve fit, the sum of the squared residuals of all points to the synthetic lightcurve is calculated. The solution with the smallest sum is best. We define 
\begin{equation}
\sigma_{\text{fit}}(x)=\sqrt{\frac{n}{n-m}\frac{1}{\sum^n_{\nu=1}\omega_{\nu}}\sum^n_{\nu=1}\omega_{\nu}d^2_{\nu}(x)}
\end{equation}
as an indicator for the quality of the fit, where $d_{\nu}$ are the residuals, the difference between observational data and the calculated model and $\omega_{\nu}$ are weights that can be used to increasingly consider several data points, i.e. the bottom of the primary eclipse, m is the number of start parameters for the simplex algorithm and n is the number of data points.\\ It was found that there is a solution for a mass ratio of q = 0.33 that reproduces the lightcurve best (see Table 1).
The reflection effect is dominating the shape of the lightcurve between the eclipses. No sign of ellipsoidal variations was found.

The solution requires an albedo for the companion that exceeds 1 (A2 =1.8), which means the concept of pure reflection is inadequate and reprocessing of radiation in the atmosphere of the companion has to be taken into account, which is beyond the scope of this work. An albedo exceeding unity has been found for other HW Vir systems as well \cite{hwviralbedo}. No reasonable fits for values of A2 =1 or less were found. Lightcurves at other wavelengths are required to better constrain the secondary's albedo. \newline
As the reduced $\chi^2$ for the solution is well above 1, the question arises if the errors in the photometric measurements are estimated well. If so, this could be a hint for unresolved pulsations superposed on the eclipse curve.

%______________________________________________________________

\section{Summary}
The star PTF1J072456$+$125301 was found to be an eclipsing binary with an orbital period of 0.09980(25)\,d or 2\,h 23.71\,min showing a strong reflection effect typical for HW Vir Stars. We obtained time series with the Palomar P60 telescope and spectroscopic data with the P200 telescope and performed an analysis of the light- and radialvelocity curves in order to pin down the systems parameters. An  quantitative spectral analysis of the co-added spectrum allowed us to determine the atmospheric parameters of the sdB. The effective temperature of T$_{\rm eff}$ = 39000$\pm$ 350\,K, log(g) = 5.74$\pm$0.08 and a helium abundance of log N$_{\rm He/H}$ = -2.02$\pm$0.07 are typical for sdB stars and place it on or near the Extreme Horizontal Branch.\newline
From the radial velocity curve we derive the mass ratio of 0.327$\pm$0.033 by adopting the canonical sdB mass (0.47\,M$_{\odot}$) for the primary. The analysis of the light curve results reveal the mass, radius and temperature of the companion of M$_{\text{comp}}$=0.155$\pm$0.020\,M$_{\odot}$, R$_{\text{comp}}$=0.165$\pm$0.015\,R$_{\odot}$ and T$_{\rm eff}$=3300$\pm$300\,K, indication that the companion is a late M dwarf.

Further observations of this interesting system should be aimed at gaining spectra with better quality. High resolution spectra may reveal lines of the companion and would allow to determine the mass ratio directly from the spectra and check if there are any discrepancies with our lightcurve-based solution.

Most important would be high-quality light curves (short cadence, high signal to noise ratio), preferably in several wavebands, to search for short-period, multi-periodic light variations caused by pulsations. Their detection would make PTF0724 an ideal target for asteroseismolgy similar to the enigmatic NY Vir \citep{vuckovic07,vuckovic09}, providing two independent tools of mass determination to be combined \citep{grootel13}. 

%______________________________________________________________
\begin{acknowledgements}
Based on observations collected at the Palomar Observatory with the 200'' Hale Telescope, operated by the California institute of Technology, its divisions Caltech Optical Observations, the Jet Propulsion Laboratory (operated for NASA) and Cornell University.
Based on observations of the Palomar Transient Factory. The Palomar Transient Factory is a scientific collaboration between the California Institute of Technology, Columbia University, Las Cumbres Observatory, the Lawrence Berkely National Laboratory, the National Energy Research Scientific Computing Center, the University of Oxford, and the Weizmann Institute of Science.\newline
T.Kupfer acknowledges support by the Netherlands Research School of Astronomy (NOVA). 
V. Schaffenroth is supported by the Deutsches Zentrum für Luft- und Raumfahrt (grant 50 OR 1110).
This research has made use of ISIS functions provided by ECAP/Remeis Obervatory.
\end{acknowledgements}

\bibliographystyle{aa}

\begin{thebibliography}{}
\bibitem[(Bertin \& Arnouts 1996)]{Bertin} Bertin, E., \& Arnouts, S. 1996, A\&AS, 117, 393
\bibitem[(Brown et al. 1997)]{brown97}Brown, T.M., Ferguson, H.C., Davidsen, A.F., \& Dorman, B., 1997, ApJ 482, 685
\bibitem[(Charpinet et al. 2007)]{charpinet07}Charpinet, S. Fontaine, G., Brassard, et al., 2007, Commun. Asteroseismol. 150, 241  
\bibitem[(Cenko et al. 2006)]{cenko06} Cenko, S. B., Fox, D. B., Moon, D.-S., et al. 2006, PASP, 118, 1396
\bibitem[(Copperwheat et al. 2011)]{copperwheat} Copperwheat, C. M., Morales-Rueda, L., Marsh, T. R., Maxted, P. F. L., Heber, U., 2011MNRAS.415.1381C
\bibitem[(Dorman et al. 1993)]{dorman}Dorman, B.,Rood,R.T., \& O'Connell, R.W. 1993, ApJ, 419,596
\bibitem[(Drechsel et al. 1995)]{drechsel95} Drechsel, H., Haas, S., Lorenz, R., \& Gayler, S. 1995, A\&A, 294, 723
\bibitem[(Drechsel et al. 2001)]{hs0705}Drechsel, H., Heber, U., Napiwotzki, R., et al. 2001, A\&A, 379, 893D
\bibitem[(Ferguson et al. 1984)]{ferguson84} Ferguson, D.~H.,
Green, R.~F., \& Liebert, J.\ 1984, \apj, 287, 320
\bibitem[(Ferraro et al. 1997)]{ferraro}Ferraro, F., Paltrinieri, B., Fusi Pecci, F., Cacciari, C., Dorman, B., \&Rood, R. 1997, ApJ 184, L 145
\bibitem[(Fontaine et al. 2012)]{fontaine12} Fontaine, G., Brassard, P., Charpinet, S., Green, E.M., Randall, S.K., Van Grootel, V. 2012A\&A...539A..12F
\bibitem[(Han et al. 2002)]{han02} Han, Z., Podsiadlowski, Ph., Maxted, P. F. L., Marsh, T. R., \& Ivanova, N. 2002, MNRAS, 336, 449
\bibitem[(Han et al. 2003)]{han03} Han, Z., Podsiadlowski, Ph., Maxted, P. F. L., \& Marsh, T. R. 2003, MNRAS, 341, 669
\bibitem[(Heber et al. 2000)]{heber2000}Heber, U., Reid, I.N., \& Werner, K. 2000, A\&A, 363, 198
\bibitem[(Heber 2009)]{heber09} Heber, U. 2009, ARAA, 47, 211
\bibitem[(Hirsch 2009)]{hirsch}Hirsch, H. 2009, Phd thesis, Friedrich Alexander Universität Erlangen Nürnberg
\bibitem[(Honeycut 1992)]{honeycut92} Honeycutt, R. K. 1992, PASP, 104, 435
\bibitem[(Horne 1986)]{horne86} Horne, K. 1986, 1986PASP..98..609H
\bibitem[(Kilkenny et al. 1978)]{kilkenny}Kilkenny, D., Hilditch, R. W., \&Penfold, J. E. 1978, MNRAS, 183, 523
\bibitem[(Kupfer et al. 2015)]{kupfer14} Kupfer, T., Geier, S., Heber, U., {\O}stensen, R. H., Barlow, B. N., Maxted, P. F. L., Heuser, C., Schaffenroth, V. G\"ansicke, B. T., 2015, 2015A\&A...576A..44K
\bibitem[(Law et al. 2009)]{law}Law, N.M. , 2009, 2009PASP..121.1395L
\bibitem[(Levenberg 1944)]{levenberg} Levenberg K., Quarterly of Applied Mathematics 2: 164–168.
\bibitem[(Levitan et al. 2011)]{levitan11} Levitan, D., Fulton, B.J., Groot, P.J., Kulkarni, S.R., Ofek, E.O., Prince, T.A., Shporer, A., Bloom, J.S., Bradley Cenko, S., Kasliwal, M.M., Law, N.M., Nugent, P.E., Poznanski, D., Quimby, R.M., Horesh, A., Sesar, B., Sternbarg, A., 2011, ApJ 739:68
\bibitem[(Lucy 1967)]{lucy}Lucy, L. B. 1967, Zeitschrift für Astrophysik, 65, 89
\bibitem[(Marsh 1989)]{marsh89} Marsh, T. R. 1989, 1989PASP..101.1032M
\bibitem[(Maxted et al. 2001)]{maxted01} Maxted, P. F. L., Heber, U., Marsh, T. R., \& North, R. C. 2001, MNRAS, 326, 1391
\bibitem[(Menzies et al. 1986)]{menzies} Menzies, J. W., Marang, F., 1986IAUS..118..305M
\bibitem[(Naylor 1998)]{naylor}Naylor, T. 1998, MNRAS, 296, 339
\bibitem[(Ofek et al. 2011)]{ofek11} Ofek, E. O., Frail, D. A., Breslauer, B., et al. 2011, 2011ApJ...740...65O
\bibitem[(Oke et al. 1982)]{oke82}Oke J. B., Gunn J. E., 1982, PASP, 94, 586 
\bibitem[(Schaffenroth et al. 2013)]{hwviralbedo} Schaffenroth, V., Geier, S., Drechsel, H., Heber, U., Wils, P, {\O}stensen, R. H., Maxted, P. F. L., di Scala, G., 2013A\&A 553A 18S
\bibitem[(Schaffenroth et al. (2014)]{schaffenroth14a} Schaffenroth, V., Geier, S., Heber, U., et al.\ 2014a, A\&A, 564, AA98 
\bibitem[Schaffenroth et al. (2014b)]{schaffenroth14b} Schaffenroth, V., 
Geier, S., Barbu-Barna, I., et al.\ 2014b, 6$^{\rm th}$ Meeting on Hot Subdwarf Stars 
and Related Objects, 481, 253 
\bibitem[(Schaffenroth et al. 2013)]{asas}Schaffenroth, V., Geier, S., Drechsel, H. et al. 2013, A\&A, 553, A18
\bibitem[Stark \& Wade(2003)]{stark03} Stark, M.~A., \& Wade, R.~A.\ 2003, \aj, 126, 1455
\bibitem[Van Grootel et 
al.(2013)]{grootel13} Van Grootel, V., Charpinet, S., Brassard, P., Fontaine, G., \& Green, E.~M.\ 2013, A\&A, 553, AA97
\bibitem[(Van Grootel et al. 2014)]{grootel} Van Grootel, V., Charpinet, S., Fontaine, G., Brassard, P., Green, E., 2014, ASPC..481..229V
\bibitem[(von Zeipel 1924)]{zeipel}von Zeipel, H.1924, MNRAS, 84, 665
\bibitem[(Vos et al. 2013)]{vos}Vos, J., \O stensen, R.H., Nemeth, P., Green, E.M., Heber, U., Van Winckel, H., 2013 A\&A, 559,A54
\bibitem[Vu{\v c}kovi{\'c} et 
al.(2007)]{vuckovic07} Vu{\v c}kovi{\'c}, M., Aerts, C., {\"O}stensen, R., et al.\ 2007, A\&A, 471, 605 
\bibitem[Vu{\v c}kovi{\'c} et 
al.(2009)]{vuckovic09} Vu{\v c}kovi{\'c}, M., {\O}stensen, R.~H., Aerts, C., et al.\ 2009, A\&A, 505, 239
\bibitem[(Wade \& Rucinski 1985)]{wade85}Wade, R.A., Rucinski, S.M. 1985, A\&A 60, 471
\bibitem[(Wilson \& Devinney 1971)]{wilson}Wilson, R.E. \& E.J. 1971, APJ, 166, 605




   
\end{thebibliography}

\Online
\begin{appendix}
\setcounter{table}{0}
\begin{table}[b]
\caption{Radial velocities with errors of PTF0724}
\label{rv_measure}
\centering
\begin{tabular}{lr}
\begin{tabular}{c|cc|c|c}
	
		HJD & \multicolumn{2}{c|}{RV in km/s}&used lines&S/N (460\,nm to 465\,nm)\\\toprule
2455477.930387&-113&$\pm15.6$&H$\beta$, H$\gamma$, H$\delta$, H$\epsilon$&11.5\\
2455477.933084&-82&$\pm23$&H$\beta$, H$\gamma$, H$\epsilon$&10.7\\
2455477.935414&-61&$\pm19.8$&H$\beta$, H$\gamma$, H$\delta$&9.5\\
2455477.937745&-105&$\pm13.2$&H$\beta$, H$\gamma$, H$\delta$, H$\epsilon$&10.8\\
2455477.940076&-116&$\pm29.5$&H$\beta$, H$\gamma$, H$\delta$, H$\epsilon$&11.3\\
2455477.942406&-86&$\pm27.1$&H$\beta$, H$\gamma$, H$\delta$, H$\epsilon$&11.3\\
2455477.944737&-95&$\pm8.2$&H$\beta$, H$\delta$, H$\epsilon$&10.0\\
2455477.949969&-86&$\pm56.8$&H$\beta$, H$\delta$, H$\epsilon$&10.4\\
2455477.952300&-151&$\pm6.9$&H$\beta$, H$\gamma$, H$\delta$, H$\epsilon$&9.9\\
2455477.954630&-99&$\pm44.8$&H$\beta$, H$\gamma$, H$\delta$, H$\epsilon$&10.0\\
2455477.956961&-85&$\pm22.4$&H$\beta$, H$\gamma$, H$\delta$, H$\epsilon$&11.7\\
2455477.959292&-96&$\pm40.3$&H$\beta$, H$\delta$, H$\epsilon$&9.8\\
2455477.961622&-60&$\pm12.2$&H$\beta$, H$\gamma$, H$\delta$, H$\epsilon$&11.4\\
2455477.963953&-37&$\pm37.1$&H$\beta$, H$\gamma$, H$\delta$, H$\epsilon$&13.7\\
2455477.966283&-74&$\pm29.2$&H$\beta$, H$\gamma$, H$\delta$, H$\epsilon$&13.7\\
2455477.968614&-131&$\pm48.8$&H$\beta$, H$\gamma$, H$\delta$, H$\epsilon$&11.8\\
2455477.970944&-67&$\pm26.8$&H$\beta$, H$\gamma$, H$\delta$, H$\epsilon$&13.0\\
2455477.973524&-53&$\pm14.3$&H$\beta$, H$\gamma$, H$\delta$, H$\epsilon$&13.2\\
2455477.980647&-16&$\pm36.4$&H$\beta$, H$\gamma$, H$\epsilon$&14.7\\
2455477.982977&-9&$\pm18.5$&H$\beta$, H$\delta$, H$\epsilon$&11.8\\
2455477.985308&+53&$\pm3.5$&H$\gamma$, H$\delta$&13.0\\
2455477.988790&+88&$\pm17.9$&H$\beta$, H$\gamma$, H$\delta$&11.9\\
2455477.991121&+32&$\pm15.2$&H$\beta$, H$\gamma$, H$\delta$, H$\epsilon$&10.2\\
2455477.993452&+149&$\pm19.9$&H$\beta$, H$\gamma$, H$\delta$, H$\epsilon$&12.5\\
2455477.995782&+36&$\pm13.8$&H$\beta$, H$\gamma$, H$\delta$, H$\epsilon$&10.0\\
2455477.998113&+73&$\pm20.5$&H$\beta$, H$\gamma$, H$\delta$, H$\epsilon$&11.2\\
2455478.000443&+88&$\pm22.4$&H$\beta$, H$\gamma$, H$\delta$, H$\epsilon$&13.4\\
2455478.002774&+49&$\pm33.5$&H$\beta$, H$\gamma$, H$\delta$, H$\epsilon$&13.2\\
2455478.005105&+48&$\pm22$&H$\beta$, H$\gamma$, H$\delta$&12.3\\
2455478.007435&+34&$\pm30.8$&H$\beta$, H$\gamma$, H$\delta$&11.8\\
2455478.009766&+82&$\pm20$&H$\beta$, H$\gamma$, H$\delta$, H$\epsilon$&13.4\\
2455478.014909&+23&$\pm13.8$&H$\beta$, H$\gamma$, H$\delta$, H$\epsilon$&13.7\\
2455478.017240&-14&$\pm17.2$&H$\beta$, H$\gamma$, H$\delta$, H$\epsilon$&11.2\\
2455478.019727&+4&$\pm27$&H$\gamma$, H$\delta$, H$\epsilon$&11.7\\
2455478.022058&-43&$\pm16.3$&H$\beta$, H$\gamma$, H$\delta$, H$\epsilon$&6.8\\
2455478.024389&-91&$\pm27.9$&H$\delta$, H$\epsilon$&4.8\\
2455478.027550&-42&$\pm23.2$&H$\gamma$, H$\delta$, H$\epsilon$&6.5\\
2455478.029881&-25&$\pm46.1$&H$\gamma$, H$\delta$, H$\epsilon$&9.0
\end{tabular}
\end{tabular}
\end{table}

\begin{table}[!h]
\caption{Beginning and End-time of the photometric observations}
\begin{tabular}{|c|c|c|c|c|}
\hline
Date&Begin (UTC)&End (UTC)&median seeing&\# of exposures\\
\hline
09.04.2010&03:10:39&06:21:59&1.61''&180\\
08.10.2010&10:01:26&12:54:16&1.16''&164\\
09.10.2010&09:59:02&12:53:53&1.64''&165\\
10.10.2010&09:56:52&12:54:47&1.83''&167\\
11.10.2010&09:50:11&12:07:28& 1.41''&130\\
13.10.2010&09:50:50&12:56:54&1.41''&176\\
15.10.2010&09:33:18&12:46:25&1.22''&180\\
\hline
\end{tabular}
\end{table}

\end{appendix}

%-------------------------------------------------------------------

\end{document}